\documentstyle[preprint,aps]{revtex}
\tightenlines

\newtheorem{theorem}{Theorem}
\newtheorem{acknowledgement}[theorem]{Acknowledgement}

\begin{document}
\title{New Singular and Nonsingular Colliding Wave Solutions in
Einstein-Maxwell-Scalar Theory.}
\author{Ozay Gurtug, Mustafa Halilsoy and Izzet Sakalli}
\address{Department of Physics, Eastern Mediterranean University\\
G. Magusa, North Cyprus, Mersin 10 - Turkey\\
email: ozay.gurtug@emu.edu.tr\\
}
\maketitle

\begin{abstract}
{\small A technique is given to generate coupled scalar field solutions in
colliding Einstein-Maxwell (EM) waves. By employing the Bell - Szekeres
solution as seed and depending on the chosen scalar field, it is possible to
construct nonsingular solutions. If the original EM solution is already
singular, addition of scalar fields does not make the physics any better. In
particular, scalar field solution that is transformable to spherical
symmetry is plagued with singularities.}
\end{abstract}

\section{Introduction}

Colliding plane waves (CPWs) in general relativity is known to result either
in an all encompassing spacelike singularity or an extendible Cauchy horizon
(CH)[1-3]. (To be more precise, this classification must be suplemented by
the rare class of null singularities). Within this context colliding
Einstein-Maxwell (CEM) waves is studied in greater detail [1,4,5,6]. While
the collision of scalar waves is also known to a certain extend [7,8], the
collision of Einstein -Maxwell-Scalar (CEMS) waves has not been investigated
in detail. Even dilaton and axion have attracted interest enough to
necessiate in retrospect study of scalar fields in its own right. By scalar
field it is implied a massless scalar field with minimal coupling, which is
the simplest type among existing fields to couple gravity and
electromagnetic (em) fields. In recent decade the effect of such a massless
scalar field, either in collapse under its own gravity or under the effect
of a black hole both have been among popular topics in general relativity
[9,10]. Minimal coupling to gravity makes a scalar field an indespensable
test field for the perturbation studies. To test the stability of a CH
formed in CEM waves we construct CEMS solutions and check whether CH remains
as CH or transforms \ into a spacetime singularity. This involves the back
reaction effect of the scalar field and must be considered as a stronger
case compared to any perturbation analysis. Being inspired by the richness
of the CPW spacetimes some researchers investigated mini black hole
formation in a laboratory by colliding highly energetic particles [11] that
can be considered analogues model of CPWs. All these aspects motivated us to
inspect the scalar field effects when coupled in collision with gravity and
em fields.

In this paper we consider first the CEM solution of Bell - Szekeres (BS) [4]
as seed and construct by the $M-shift$ method, scalar field solutions
coupled with it. As a second example we consider the solution found by
Hogan, Barrabes and Bressange (HBB) [12]. This solution represents an
example of colliding impulsive gravitational wave with a wavepacket
consisting of superposed impulsive gravitational wave with an em shock wave.
Since this is already a singular solution the added scalar field serves only
to modify the colliding wavepackets. Our final example is a scalar field
solution that is transformable to the spherically symmetric Penney solution
[13].

The paper is organized as follows. In section II we introduce and prove the $%
M-shift$ method for generating scalar field solutions. In sections III and
IV we apply our method to the BS and HBB metrics, respectively. Section V
contains a singular solution from the outset because it has been considered
isometric to the spherically symmetric geometry. We conclude the paper with
discussion in section VI.

\section{A Method For Generating CEMS Solutions From Any Known CEM Solution.}

In this section we give a simple proof that when given any solution in CEM
theory we can generate \ a class of CEMS solutions with prescribed
properties. A similar technique had been used long time ago in the
Weyl-Papapetrou form of stationary axially symmetric gravitational fields
[14]. The similarity between this form of the metric and the \ metric of
CPWs serves to extend the technique automatically to the latter.

The metric of CPWs in general is given by 
\begin{equation}
ds^{2}=2e^{-M}dudv-e^{-U}[(e^{V}dx^{2}+e^{-V}dy^{2})\cosh W-2\sinh W\,dxdy]
\end{equation}

in which it is understood that all metric functions are at most functions of
the null coordinates $u$ and $v$. We also add that whatever we prove for the
EM system a reduced version is also valid for the vacuum CPWs. By choosing
an em potential one form

\begin{equation}
\tilde{A}=\tilde{A_{\mu }}dx^{\mu }=Adx+Bdy
\end{equation}

where $A$ and $B$ are potential functions to be determined and a scalar
field $\phi $, the CEMS system can be generated from the Lagrangian density 
\begin{eqnarray}
L &=&e^{-U}\left(
M_{u}U_{v}+M_{v}U_{u}+U_{u}U_{v}-W_{u}W_{v}-V_{u}V_{v}cosh^{2}W-4\phi
_{u}\phi _{v}\right)  \nonumber \\
&&-2k\left[ \left( B_{u}B_{v}e^{V}+A_{u}A_{v}e^{-V}\right) coshW\right. 
\nonumber \\
&&\left. \hspace{2.5cm}+\left( A_{u}B_{v}+A_{v}B_{u}\right) sinhW\right]
\end{eqnarray}
\ \ 

The constant $k$ denotes a coupling constant which can be fixed as unity.
Variational principle of the action $S$ defined by (and suppressing the $x,y$
coordinates)

\begin{equation}
S=\int Ldudv
\end{equation}

yields the following CEMS equations 
\begin{eqnarray}
U_{uv} &=&U_{u}U_{v} \\
2M_{uv} &=&-U_{u}U_{v}+W_{u}W_{v}+V_{u}V_{v}cosh^{2}W+4\phi _{u}\phi _{v} \\
2V_{uv} &=&U_{v}V_{u}+U_{u}V_{v}-2\left( V_{u}W_{v}+V_{v}W_{u}\right) tanhW 
\nonumber \\
&&-2ksechW\left( \bar{\Phi _{0}}\Phi _{2}+\bar{\Phi _{2}}\Phi _{0}\right) \\
2W_{uv} &=&U_{v}W_{u}+U_{u}W_{v}+2V_{u}V_{v}coshWsinhW  \nonumber \\
&&+2ki\left( \bar{\Phi _{0}}\Phi _{2}-\bar{\Phi _{2}}\Phi _{0}\right) \\
2\phi _{uv} &=&U_{v}\phi _{u}+U_{u}\phi _{v} \\
2A_{uv} &=&V_{v}A_{u}+V_{u}A_{v}-tanhW\left( W_{v}A_{u}+W_{u}A_{v}\right) 
\nonumber \\
&&-e^{V}\left[ 2B_{uv}tanhW+W_{u}B_{v}+W_{v}B_{u}\right] \\
2B_{uv} &=&-V_{v}B_{u}-V_{u}B_{v}-tanhW\left( W_{v}B_{u}+W_{u}B_{v}\right) 
\nonumber \\
&&-e^{V}\left[ 2A_{uv}tanhW+W_{u}A_{v}+W_{v}A_{u}\right]
\end{eqnarray}
Here $\Phi _{0}$ and $\Phi _{2}$ are the Newman-Penrose spinors for em
fields defined by 
\begin{eqnarray}
\Phi _{2} &=&\frac{e^{U/2}}{\sqrt{2}}\left[ e^{-V/2}\left( isinh\frac{W}{2}%
-cosh\frac{W}{2}\right) A_{u}\right.  \nonumber \\
&&\left. +e^{V/2}\left( icosh\frac{W}{2}-sinh\frac{W}{2}\right) B_{u}\right]
\nonumber \\
\Phi _{0} &=&\frac{e^{U/2}}{\sqrt{2}}\left[ e^{-V/2}\left( isinh\frac{W}{2}%
+cosh\frac{W}{2}\right) A_{v}\right.  \nonumber \\
&&\left. +e^{V/2}\left( icosh\frac{W}{2}+sinh\frac{W}{2}\right) B_{v}\right]
\end{eqnarray}
The remaining equations corresponding to $R_{uu}=-T_{uu}$ and $%
R_{vv}=-T_{vv} $ which do not follow from the variational principle, namely 
\begin{eqnarray}
2U_{uu}-U_{u}^{2}+2M_{u}U_{u} &=&W_{u}^{2}+V_{u}^{2}cosh^{2}W+4\phi
_{u}^{2}+4k\Phi _{2}\bar{\Phi _{2}}  \nonumber \\
2U_{vv}-U_{v}^{2}+2M_{v}U_{v} &=&W_{v}^{2}+V_{v}^{2}cosh^{2}W+4\phi
_{v}^{2}+4k\Phi _{0}\bar{\Phi _{0}}
\end{eqnarray}
are automatically satisfied by virtue of the other equations. Thus the
foregoing sets of equations (5 -13) give the complete set of CEMS equations.
The equation (13) actually are the guiding equations for us to state the
following

\begin{theorem}
\bigskip Given an EM metric satisfying the above equations (5-13) with zero
scalar field ($\phi =0$), we can generate solutions with $\phi \neq 0$ \ by
making a shift in the metric function $M$ (i.e. the $M$ - shift) in
accordance with
\end{theorem}

\begin{equation}
M\longrightarrow \widetilde{M}=M+\Gamma
\end{equation}
where the shift function $\Gamma $ must satisfy 
\begin{eqnarray}
\Gamma _{u}U_{u} &=&2\phi _{u}^{2}  \nonumber \\
\Gamma _{v}U_{v} &=&2\phi _{v}^{2}
\end{eqnarray}

\bigskip {\bf Proof:} We observe easily from the pair of eqs.(13) that
substitution of $M\longrightarrow \widetilde{M}$, and by virtue of (15)
cancels the scalar field from both sides. Further, the integrability of
(15), i.e. $\Gamma _{uv}=\Gamma _{vu}$ imposes the scalar field equation 
\begin{equation}
2\phi _{uv}-U_{v}\phi _{u}-U_{u}\phi _{v}=0
\end{equation}
as a constraint condition. This leaves no trace of the scalar field which
completes the proof. We can justify the $M$ $-shift$ also by employing \ the
action principle which turns out to yield total divergences. In conclusion,
if we have a CEM solution consisting of ($U,M,V,W,A,B$) then we obtain a
CEMS solution expressed by ($U,M+\Gamma ,V,W,A,B,\phi $) where $\Gamma $ is
obtained as a line integral 
\begin{equation}
\Gamma =2\int \frac{\phi _{u}^{2}}{U_{u}}\,du+2\int \frac{\phi _{v}^{2}}{%
U_{v}}\,dv
\end{equation}
In this \ technique em potentials and the metric functions $U,V$ and $W$
remain unchanged. Also in practice, the line integral (17) is (except in
very particular cases) of little use. Transforming from the null coordinates
($u,v$) to new types of coordinates ($\tau ,\sigma $), however, we get more
advantage toward solutions in closed form. One such useful set of
coordinates is defined by

\begin{eqnarray}
\tau &=&u\sqrt{1-v^{2}}+v\sqrt{1-u^{2}} \\
\sigma &=&u\sqrt{1-v^{2}}-v\sqrt{1-u^{2}}  \nonumber
\end{eqnarray}

which transforms the relavant part of the metric as

\begin{equation}
\frac{d\tau ^{2}}{\Delta }-\frac{d\sigma ^{2}}{\delta }=\frac{4dudv}{\sqrt{%
1-u^{2}}\sqrt{1-v^{2}}}
\end{equation}

with 
\begin{eqnarray}
\Delta &=&1-\tau ^{2} \\
\delta &=&1-\sigma ^{2}  \nonumber
\end{eqnarray}

In terms of these new coordinates the scalar field equation (16) takes the
form

\begin{equation}
\left( \Delta \phi _{\tau }\right) _{\tau }-\left( \delta \phi _{\sigma
}\right) _{\sigma }=0
\end{equation}

while the $\Gamma $ eq.s (15) become

\begin{eqnarray}
(\tau ^{2}-\sigma ^{2})\Gamma _{\tau } &=&2\Delta \delta \left( \tau \phi
_{\tau }^{2}+\frac{\tau \delta }{\Delta }\phi _{\sigma }^{2}-2\sigma \phi
_{\tau }\phi _{\sigma }\right)  \nonumber \\
&&  \nonumber \\
(\sigma ^{2}-\tau ^{2})\Gamma _{\sigma } &=&2\Delta \delta \left( \sigma
\phi _{\sigma }^{2}+\frac{\sigma \Delta }{\delta }\phi _{\tau }^{2}-2\tau
\phi _{\tau }\phi _{\sigma }\right)
\end{eqnarray}
The advantage we have obtained by this change of coordinates is that the
scalar field equation (21) admits an infinite class of seperable solutions
which were not so obvious in the original null coordinates.

A general class of separable solution for the scalar field $\phi $ is given
by [1] 
\begin{equation}
\phi (\tau ,\sigma )=\sum_{n}\{a_{n}P_{n}(\tau )P_{n}(\sigma
)+b_{n}Q_{n}(\tau )Q_{n}(\sigma )+c_{n}P_{n}(\tau )Q_{n}(\sigma
)+d_{n}P_{n}(\sigma )Q_{n}(\tau )\}
\end{equation}
Where $P$ and $Q$ are the Legendre functions of the first and second kind,
respectively and $a_{n},b_{n},c_{n}$ and $d_{n}$ are arbitrary constants.
Although this expression for $\phi (\tau ,\sigma )$ together with the
integrals (22) solve the scalar field problem mathematically, we must impose
also some physical conditions. The Cauchy data to be imposed as physical
input must be acceptable. This discards from the outset any diverging
solutions for the scalar field or the metric function $e^{-\Gamma }$ which
has unacceptable incoming limits. The $M$ $-shift$ technique changes the
Weyl scalars while keeps the em energy momentum of the CEM problem
unchanged. Under the light of all these considerations we present examples
of scalar field solutions to some important CEM solutions.

\section{Bell - Szekeres Solution Coupled with Scalar Fields.}

CEM waves with constant profiles is known as the BS solution given by the
line element

\begin{equation}
ds^{2}=2dudv-\cos ^{2}\left( au+bv\right) dx^{2}-\cos ^{2}\left(
au-bv\right) dy^{2}
\end{equation}
\newline
\ where ($a,b)$ are the constants of em fields. In the new coordinates

\begin{eqnarray}
\tau &=&\sin \left( au+bv\right) \\
\sigma &=&\sin \left( au-bv\right)  \nonumber
\end{eqnarray}

this line element takes the form

\begin{equation}
ds^{2}=\frac{1}{2ab}\left( \frac{d\tau ^{2}}{\Delta }-\frac{d\sigma ^{2}}{%
\delta }\right) -\left( \Delta dx\right) ^{2}-\left( \delta dy\right) ^{2}
\end{equation}

We note that these $\left( \tau ,\sigma \right) $ coordinates can be
obtained from the ones of previous section by letting $u\longrightarrow \sin
\left( au\right) $ and $v\longrightarrow \sin \left( bv\right) $. In the BS
metric we have the case that $M=0$, therefore by the $M$ $-shift$ we obtain

\begin{equation}
ds^{2}=\frac{e^{-\Gamma }}{2ab}\left( \frac{d\tau ^{2}}{\Delta }-\frac{%
d\sigma ^{2}}{\delta }\right) -\left( \Delta dx\right) ^{2}-\left( \delta
dy\right) ^{2}
\end{equation}

Now by choosing the scalar field

\begin{equation}
\phi \left( \tau ,\sigma \right) =\alpha \tau \sigma +\frac{1}{4}\beta
\left( 3\tau ^{2}-1\right) \left( 3\sigma ^{2}-1\right)
\end{equation}

as a solution of (21), with $\left( \alpha ,\beta \right) $ arbitrary
constants, it enables us to integrate $\Gamma $ from (22) with the result

\begin{eqnarray}
\Gamma &=&\alpha ^{2}\tau ^{2}+\frac{9}{4}\beta ^{2}\tau ^{2}(1-\frac{\tau
^{2}}{2})-6\alpha \beta \tau \sigma \Delta \delta +  \nonumber \\
&&  \nonumber \\
&&\frac{\Delta }{4}\left\{ \frac{9}{2}\beta ^{2}\sigma ^{2}(9\tau
^{2}-1)+\sigma ^{2}\left( 4\alpha ^{2}+9\beta ^{2}-45\beta ^{2}\tau
^{2}\right) \right\}
\end{eqnarray}

This function is well defined and finite as the CH at $\tau =1$ is
approached. For $\tau \longrightarrow 1,$ we have

\begin{equation}
\Gamma \left( \tau \longrightarrow 1\right) =\alpha ^{2}+\frac{9}{8}\beta
^{2}
\end{equation}

This constitutes a non-singular extension of the BS solution in the presence
of a scalar field. We consider the simpler case with $\beta =0$, which has
the following scale invariant Weyl scalars

\begin{eqnarray}
\Psi _{2}^{\left( 0\right) } &=&-\alpha ^{2}ab\theta \left( u\right) \theta
\left( v\right) \sin \left( 2au\right) \sin \left( 2bv\right) \\
&&  \nonumber \\
\Psi _{4}^{\left( 0\right) } &=&a^{2}\theta \left( v\right) \left[ \delta
\left( au\right) \tan \left( bv\right) +\alpha ^{2}\theta \left( u\right)
\sin \left( 2au\right) \sin \left( 2bv\right) \right]  \nonumber \\
&&  \nonumber \\
\Psi _{0}^{\left( 0\right) } &=&b^{2}\theta \left( u\right) \left[ \delta
\left( bv\right) \tan \left( au\right) +\alpha ^{2}\theta \left( v\right)
\sin \left( 2au\right) \sin \left( 2bv\right) \right]  \nonumber
\end{eqnarray}

The Ricci components are also given (in the Newman - Penrose formalism) by

\begin{eqnarray}
\Phi _{02}^{\left( 0\right) } &=&-ab\theta \left( u\right) \theta \left(
v\right) \\
&&  \nonumber \\
\Phi _{22}^{\left( 0\right) } &=&a^{2}\theta \left( u\right) \left[ 1+\alpha
^{2}\sin ^{2}\left( 2au\right) \right]  \nonumber \\
&&  \nonumber \\
\Phi _{00}^{\left( 0\right) } &=&b^{2}\theta \left( v\right) \left[ 1+\alpha
^{2}\sin ^{2}\left( 2bv\right) \right]  \nonumber \\
&&  \nonumber \\
\Phi _{11}^{\left( 0\right) } &=&-3\Lambda ^{\left( 0\right) }=-\frac{1}{2}%
\alpha ^{2}ab\theta \left( u\right) \theta \left( v\right) \sin \left(
2au\right) \sin \left( 2bv\right)  \nonumber
\end{eqnarray}

It is seen that with the exception of the distributional singularities on $%
\left( u=0,bv=\pi /2\right) $ and $\left( v=0,au=\pi /2\right) $ to the
future of the collision the metric is free of singularities.

We note that this scalar field extension of the BS metric also applies to
its cross-polarized version easily. Since this is an exact back-reaction
solution to the CEMS fields it provides an example that scalar field
perturbations need not transform the CH into singularity. Different scalar
fields, however, may not preserve the regularity of the CH. Hence, it should
not be wrong to conclude that, the stability or instability of CHs against
scalar field perturbations depends crucially on the perturbing scalar field
potential.

\section{Scalar Field Extension of the HBB Solution.}

An interesting solution in the CEM waves was given by HBB which represents
collision of an impulsive gravitational wave with a wave packet consisting
of superposed impulsive gravitational wave and a shock em wave. This
solution naturally possesses both the Khan - Penrose [15] and Griffiths [16]
limits and does not belong to any known family of solutions [1]. In this
section we show that by the $M-shift$ \ we can add a scalar field to the
colliding fields to extend them into more complex wave packets.

The incoming metrics in the HBB problem are [12] 
\begin{equation}
ds^{2}=2dudv-\left( 1+ku\right) ^{2}dx^{2}-\left( 1-ku\right) ^{2}dy^{2},%
\text{(Region II)}
\end{equation}

\bigskip

\begin{equation}
ds^{2}=2dudv-\left( \cos bv+\frac{l}{b}\sin bv\right) ^{2}dx^{2}-\left( \cos
bv-\frac{l}{b}\sin bv\right) ^{2}dy^{2},\text{(Region III)}
\end{equation}

in which the null coordinates are to be multiplied with the step functions.
Here, $k$ and $l$ are the impulsive gravitational constants while $b$
represents the em constant. We note that our coordinate $v$ (Region III) is
different from the one employed by HBB, i.e. the relation is

\begin{equation}
v\longrightarrow \frac{1}{b}\tan \left( bv\right)
\end{equation}

so that in the limit, $b\longrightarrow 0$ they coincide. The metric
functions and the em field strengths found by HBB are

\begin{eqnarray}
e^{-U} &=&F\cos ^{2}bv \\
&&  \nonumber \\
e^{V} &=&\frac{1+kuB+A\sqrt{1-B^{2}}}{1-kuB-A\sqrt{1-B^{2}}}  \nonumber \\
&&  \nonumber \\
e^{-M} &=&\frac{H^{2}}{AB\sqrt{F}}  \nonumber \\
&&  \nonumber \\
\phi _{2} &=&\frac{-kB\tan bv}{AH\sqrt{F}}  \nonumber \\
&&  \nonumber \\
\phi _{0} &=&\frac{b\left[ ku\left( \frac{l^{2}+b^{2}}{l^{2}}\right) \left(
1-B^{2}\right) ^{3/2}+AB^{3}\right] }{BH\sqrt{F}}  \nonumber
\end{eqnarray}

where the notation is

\begin{eqnarray}
F &=&A^{2}+B^{2}-1-k^{2}u^{2}\tan ^{2}bv \\
H &=&AB-ku\sqrt{1-B^{2}}  \nonumber
\end{eqnarray}

and

\begin{eqnarray}
A &=&\sqrt{1-k^{2}u^{2}} \\
B &=&\sqrt{1-\frac{l^{2}}{b^{2}}\tan ^{2}bv}  \nonumber
\end{eqnarray}

Our new coordinates appropriate for the present problem are

\begin{eqnarray}
\tau &=&B\cos bv\sqrt{1-A^{2}}+A\sqrt{1-B^{2}\ \cos ^{2}bv} \\
\sigma &=&B\cos bv\sqrt{1-A^{2}}-A\sqrt{1-B^{2}\ \cos ^{2}bv}  \nonumber
\end{eqnarray}

so that the metric function $U$ is expressed by

\begin{equation}
e^{-U}=\sqrt{1-\tau ^{2}}\sqrt{1-\sigma ^{2}}
\end{equation}

Solution of the scalar field equation (21) in the present coordinates can
easily be found. We present two particular solutions.

{\bf a)} Let 
\begin{equation}
\phi \left( \tau ,\sigma \right) =\alpha \tau \sigma
\end{equation}

where $\alpha =$ constant, and integration of $\Gamma $ function from Eq
(22) results in

\begin{equation}
\Gamma =\alpha ^{2}\left( \tau ^{2}+\sigma ^{2}-\tau ^{2}\sigma ^{2}\right)
\end{equation}

This choice of scalar field occurs from both sides of the incoming waves and
it is regular. The em field strengths remain unchanged.

{\bf b) }Let

\begin{equation}
\phi \left( \tau ,\sigma \right) =\left\{ 
\begin{array}{c}
=\beta \tanh ^{-1}\left( \frac{\tau +\sigma }{1+\tau \sigma }\right) \text{\
\ (Region IV, \ \ }u>0\text{, \ \ \ }v>0) \\ 
=0,\text{ \ \ \ (Region III, \ \ \ }u\leq 0\text{)}
\end{array}
\right.
\end{equation}

where $\beta =$ constant. The $\Gamma $ function now becomes

\begin{equation}
e^{-\Gamma }=\left[ \frac{\left( 1-\tau ^{2}\right) \left( 1-\sigma
^{2}\right) }{\left( \tau +\sigma \right) ^{4}}\right] ^{\beta ^{2}}
\end{equation}

In this particular class the scalar field exists only for $u>0$, which in
the Region II (v%
\mbox{$<$}%
0) takes the form

\begin{equation}
e^{-\Gamma }=\left( \frac{1-k^{2}u^{2}}{4k^{2}u^{2}}\right) ^{2\beta ^{2}}
\end{equation}

and is well-defined. This solitonic scalar field occurs only in Region II
and IV while in Region III there is no scalar field. Hence, Region II
contains gravity + scalar waves while Region III contains gravity +em waves.

\section{CEMS Waves Isometric to the Penney Solution.}

As another example we consider a solution for CEMS waves which is
transformable to the spherically symmetric geometry. Unlike the two previous
examples the present one has not been obtained by the $M$ $-shift$ method.
In spherically symmetric problem by the uniqueness arguments Reissner -
Nordstr\"{o}m solution is the single available black hole solution. Scalar
field extension of this metric was found long ago by Penney [13]. The result
was that inclusion of scalar field converted both horizons into spacetime
singularities which naturally destroyed the black hole property. By the same
token solution in CEMS waves that is isometric to spherically symmetric
geometry no different result other than a metric plagued with singularities
is expected.

The metric, scalar field and the em vector potential in the Region II are
given respectively as follows

\begin{eqnarray}
ds^{2} &=&Z^{2}\left( 1-u^{2}\theta (u)\right) ^{\frac{1}{2}-A}\left(
4dudv-\left( 1-u^{2}\theta (u)\right) ^{\frac{3}{2}}dx^{2}\right)
-Z^{-2}\left( 1-u^{2}\theta (u)\right) ^{A}dy^{2} \\
&&  \nonumber \\
\phi (u) &=&\frac{1}{2}\sqrt{1-A^{2}}\ln {\left| \frac{1+u\theta \left(
u\right) }{1-u\theta \left( u\right) }\right| }  \nonumber \\
&&  \nonumber \\
A_{\mu }(u) &=&2\sqrt{|ab|}\delta _{\mu }^{x}Au\theta \left( u\right) 
\nonumber
\end{eqnarray}

where $2Z(u)=a\left( 1+u\theta \left( u\right) \right) ^{A}+b\left(
1-u\theta \left( u\right) \right) ^{A}$, $\left( a,b\right) $ are the
constant em parameters, and $0\leq A\leq 1$ is the constant scalar field
parameter. Unfortunately this data has diverging energy- momentum $T_{uu}$
and Weyl scalar $\Psi _{4}$ at $u=1$. Replacing $u\longleftrightarrow v$ \ ($%
-v$ in $A_{\mu }$ ) specifies also the initial data in the incoming Region
III. The solution of these CEMS waves is

\begin{eqnarray}
ds^{2} &=&\Delta ^{1-A}Z^{2}\left( \frac{d\tau ^{2}}{\Delta }-\frac{d\sigma
^{2}}{\delta }-\delta dx^{2}\right) -\Delta ^{A}Z^{-2}dy^{2} \\
&&  \nonumber \\
\phi (\tau ) &=&\frac{1}{2}\sqrt{1-A^{2}}\ln \left| \frac{1+\tau }{1-\tau }%
\right|  \nonumber \\
&&  \nonumber \\
A_{\mu } &=&2\delta _{\mu }^{x}\sqrt{|ab|}A\sigma  \nonumber
\end{eqnarray}

\ 

where $\left( \tau ,\sigma \right) $ coordinates are as in (18) and

\begin{equation}
2Z=a(1+\tau )^{A}+b(1-\tau )^{A}
\end{equation}

We note that this solution is invariant under $A\longrightarrow -A$
therefore it is sufficient to consider the case $0\leq A\leq 1.$ As
particular limits of (47) we observe the following cases.

i) For $A=1$ (and $a=b$ ), it reduces to the well known BS solution of CEM
waves which is regular. This admits a CH at $\tau =1$ $\left( 0<\sigma
<1\right) $ and null singular points at $\tau =1$, $\sigma =\pm 1$ (i.e. $%
u=1,v=0$ and $v=1,u=0$ ).

ii) For $A=0$, it reduces to colliding Einstein - Scalar waves with a
spacelike singularity at $\tau =1$. Let us note that, it is still an open
problem to find colliding pure scalar waves without singularities.

iii) For $0<A<1$ we have an example of CEMS waves solution with a spacelike
singularity at $\tau =1.$ Further, at $\tau =1$ the metric becomes
completely degenerate, i.e. $ds^{2}=0$.

In order to see the role of the scalar field in directing the geodesics of a
particle in the interaction region we find the proper time of fall into the
singularity.

The proper time of fall into the singularity is given by

\begin{equation}
t_{0}=\int_{0}^{1}\frac{Z^{2}}{\sqrt{\delta _{1}\Delta ^{A}Z^{2}+\alpha
^{2}\Delta ^{2A-1}}}d\tau
\end{equation}

where $\alpha $ is a constant associated with a cyclic coordinate and $%
\delta _{1}=0$ (for null) or $\delta _{1}=1$ (for timelike) geodesics. We
obtain 
\begin{equation}
t_{0}=\left\{ 
\begin{array}{c}
\frac{a^{2}}{\alpha }B_{\frac{1}{2}}\left[ \frac{3}{2}-A,\frac{3}{2}+A\right]
+\frac{b^{2}}{\alpha }B_{\frac{1}{2}}\left[ \frac{3}{2}+A,\frac{3}{2}-A%
\right] +\frac{ab\pi }{8\alpha },(null-geodesics) \\ 
aB_{\frac{1}{2}}\left[ 1-\frac{A}{2},1+\frac{A}{2}\right] +bB_{\frac{1}{2}}%
\left[ 1+\frac{A}{2},1-\frac{A}{2}\right] ,(timelike-geodesics)
\end{array}
\right.
\end{equation}

in which $B_{\lambda }(\mu ,\nu )$ is an incomplete beta function defined by

\begin{eqnarray}
B_{\lambda }\left[ \mu ,\nu \right] &=&\int_{0}^{\lambda }t^{\mu
-1}(1-t)^{\nu -1}dt=\mu ^{-1}\lambda ^{\mu }F\left( \mu ,1-\nu ;\mu
+1;\lambda \right)  \nonumber \\
&&0\leq \lambda \leq 1  \nonumber \\
&&\mu ,\nu >0
\end{eqnarray}

Finally we prove the local equivalence of our metric with that of Penney
[13]. By choosing 
\[
2Z=a_{0}|1+\tau |^{A}-b_{0}|1-\tau |^{A} 
\]

and using the transformation

\bigskip 
\begin{equation}
\tau =\frac{m-r}{\sqrt{m^{2}-Q^{2}}},\hspace{0.5cm}x=\phi ,\hspace{0.5cm}y=(%
\sqrt{m^{2}-Q^{2}})t,\hspace{0.5cm}\sigma =\cos \theta
\end{equation}
with $Q^{2}=\frac{e^{2}}{A^{2}}$, where $e$ is an electric charge,
transforms our metric (47) into 
\begin{equation}
ds^{2}=e^{-\alpha }dt^{2}-e^{\alpha }dr^{2}-e^{\beta }\left( d\theta
^{2}+\sin ^{2}\theta d\phi ^{2}\right)
\end{equation}
Here we have 
\begin{eqnarray}
e^{\alpha } &=&[(r-a_{0})(r-b_{0})]^{-A}\left\{ \frac{%
b_{0}|r-a_{0}|^{A}-a_{0}|r-b_{0}|^{A}}{b_{0}-a_{0}}\right\} ^{2}  \nonumber
\\
&&  \nonumber \\
&&e^{\beta }=[(r-a_{0})(r-b_{0})]e^{\alpha }  \nonumber \\
&&  \nonumber \\
a_{0} &=&m-\sqrt{m^{2}-\frac{e^{2}}{A^{2}}}  \nonumber \\
&&  \nonumber \\
b_{0} &=&m+\sqrt{m^{2}-\frac{e^{2}}{A^{2}}}
\end{eqnarray}
Metric (53) is recognized as the solution of Penney, representing a singular
scalar field extension of the Reissner - Nordstrom geometry.

\section{Discussion.}

We presented a method that adds scalar fields to any known EM solution in
CPWs. Physically interesting case is to find solutions without
singularities. This seems possible when the background CEM metric is
singularity free in the interaction region. Any solution that is already
singular becomes worse with the addition of scalar fields. So far no
singularity free colliding pure scalar field solution has been found. In the
solution in section IV, we see dramatically how the addition of the scalar
field parameter $0<A<1$ makes spacetime singular. The $M$ - shift technique
applies equally well to any vacuum metric. The resulting superposition of
plane waves with scalar fields is equivalent to the collision of wavepackets

\begin{acknowledgement}
.We wish to thank Dr. Andrew Shoom for fruitful discussions.
\end{acknowledgement}

\end{document}